\documentclass[12pt]{article}
\usepackage{amsmath}
\usepackage{amssymb}
\usepackage{epsfig}
\let\sstl=\scriptscriptstyle
\def\Was{W\c as}
\def\Order#1{${\cal O}(#1$)}
\def\Ordpr#1{${\cal O}(#1)_{prag}$}
\def\bbe{\bar{\beta}}
\def\tbe{\tilde{\beta}}
\def\tal{\tilde{\alpha}}
\def\tom{\tilde{\omega}}
\def\half{ {1\over 2} }
\def\alf1{ {\alpha\over\pi} }

\begin{document}
 
\begin{titlepage}
 \begin{flushright}
 {\bf CERN-TH/99-222}\\
 {\bf UTHEP-98-0502 }\\
\end{flushright}
 
\begin{center}
{\Large Final-State Radiative Effects for the Exact ${\cal O}(\alpha)$
YFS\\
 Exponentiated (Un)Stable 
$W^+W^-$ Production\\ 
At and Beyond LEP2 Energies$^{\dagger}$
}
\end{center}

\vspace{2mm}
\begin{center}
  {\bf   S. Jadach$^{a,b,c}$,}
  {\bf   W. P\l{}aczek$^{d,c}$,}
  {\bf   M. Skrzypek$^{b,c}$,}
  {\bf   B.F.L. Ward$^{e,f,c}$}
  {\em and}
  {\bf   Z. W\c{a}s$^{b,c}$}

\vspace{2mm}
{\em $^a$DESY, Theory Division, D-22603 Hamburg, Germany}\\
{\em $^b$Institute of Nuclear Physics,
  ul. Kawiory 26a, 30-055 Cracow, Poland,}\\
{\em $^c$CERN, Theory Division, CH-1211 Geneva 23, Switzerland,}\\
{\em $^d$  Institute of Computer Science, Jagellonian University,\\
        ul. Nawojki 11, 30-072 Cracow, Poland,}\\
{\em $^e$Department of Physics and Astronomy,\\
  The University of Tennessee, Knoxville, Tennessee 37996-1200, USA,}\\
{\em $^f$SLAC, Stanford University, Stanford, California 94309, USA.}
\end{center}


\vspace{5mm}
\begin{center}
{\bf   Abstract}
\end{center}

We present the leading-logarithm (LL) final-state radiative effects for the  
exact ${\cal O}(\alpha)$ YFS
exponentiated (un)stable $WW$ pair production at LEP2/NLC energies
using Monte Carlo event generator methods. 
The corresponding event generator, version 1.12 of the program YFSWW3, 
wherein both Standard Model and anomalous
triple gauge-boson couplings are allowed, generates $n(\gamma)$
radiation both from the initial state and from the intermediate $W^+ W^-$
state, and generates the LL final state $W$ decay radiative effects.
Sample Monte Carlo data are given for illustration.

\vspace{10mm}
\begin{center}
{\it To be submitted to Physical Review D}
\end{center}
\vspace{10mm}
\renewcommand{\baselinestretch}{0.1}
\footnoterule
\noindent
{\footnotesize
\begin{itemize}
\item[${\dagger}$]
Work partly supported by the Polish Government
grants KBN 2P30225206 and 2P03B17210, the Maria Sk\l{}odowska-Curie
Joint Fund II PAA/DOE-97-316, and
by the US Department of Energy Contracts  DE-FG05-91ER40627
and   DE-AC03-76ER00515.
\end{itemize}
}
\begin{flushleft}
{\bf CERN-TH/99-222}\\ 
{\bf UTHEP-98-0502}\\
{\bf July, 1999}\\
\end{flushleft}

\end{titlepage}


The role of the final-state radiative (FSR) effects in the processes 
$e^+e^- \to W^+W^- +n(\gamma)\to 4f+n(\gamma)$
at and beyond LEP2
energies is of considerable interest 
for the LEP2 and NLC physics 
programs~\cite{lep2ybk:1996,frits:1998,frits:1999}.
In this paper, we evaluate for the first time the possible
interplay between the exact ${\cal O}(\alpha)$ electroweak (EW)
corrections and the leading-logarithm (LL) final-state 
radiative effects for these processes 
when the $n(\gamma)$ radiation is realized according 
to the amplitude-based Monte Carlo event generator techniques
described in Refs.~\cite{yfsww2:1996,yfsww3:1998}, wherein
infrared singularities are cancelled to all orders in $\alpha$
by using the extension to spin~1 charged particles of the
theory of Yennie, Frautschi and Suura for QED~\cite{yfs:1961}.

   The final-state radiative effects are realized in the LL 
approximation using the calculation of the program 
PHOTOS (Ref.~\cite{photos:1994}) in which a non-radiative 
final-state process is used to generate up to two photons in the
corresponding radiative process by iterating the 
structure function evolution equation%
\footnote{To be precise an ansatz is provided, which reproduces the LL terms.
It includes transverse degrees of freedom for the photon 4-momentum,  
assures coverage of the full phase space and rules of energy-momentum 
conservation.
The photons' angular distribution is chosen to reproduce exactly the one 
of the soft photon limit. See Ref.~\cite{photos:1994} for more details.
} 
for QED~\cite{strfn}.
The exact ${\cal O}(\alpha)$ YFS exponentiated final-state $W$ decay
radiative effects will be published elsewhere~\cite{elsewh}. 
In this connection, we note
that we expect the non-leading ${\cal O}(\alpha)$ and higher
order (${\cal O}(\alpha^n),~n \ge 2$) final-state radiative effects
to be small, $\sim 1\%$ in the peak reduction effect~\cite{frits:1998}
for example, even for a 
``bare trigger'' 
acceptance for
the outgoing final charged particles. This has been found by 
the authors of Ref.~\cite{frits:1998}, who analysed the effects
of final-state radiation in $Z$ decay in the naive exponentiated
(exact and LL) ${\cal O}(\alpha)$ approximation 
and who estimated the corresponding
size of the analogous effects in $W$ decay, such as $\sim 14\%$ for the total
peak reduction effect.
Indeed, more recently, the authors of Ref.~\cite{frits:1998} have
made an independent cross check on their estimates of the 
FSR line-shape effects for 
$e^+e^-\rightarrow W^+W^-\rightarrow 4f$ 
in Ref.~\cite{frits:1999}. There they present an exact ${\cal O}(\alpha)$
calculation of the process in the double-pole approximation (DPA),
wherein they retain in the pole expansion~\cite{robin:1991} of the 
complete $e^+e^-\rightarrow 4f$
amplitude, only the terms containing the double pole in the S-matrix
at the complex mass squared, 
$M^2 = M_W^2-iM_W\Gamma_W$,
where $M_W,\Gamma_W$
are the respective mass and width of the $W$ boson,
and where in the residues of the respective double poles
they project the respective ${\cal O}(\alpha)$ corrections
to an appropriate on-shell point. Henceforth, we refer to the
on-shell residue projected DPA as to the leading pole 
approximation (LPA), with more general applications in mind:
for example, in a triply resonant process, the LPA would
correspond to the triple pole terms in the respective S-matrix
element with the residues projected to an appropriate on-shell point. In this
gauge-invariant calculation, these authors find that the 
FSR peak reduction effect is $\sim 14.4\%$ for 
$W^{+(-)}\rightarrow e^{+(-)}\nu_e(\bar\nu_e)$, to be compared with
their estimate of $\sim 14\%$ in Ref.~\cite{frits:1998}.
We will compare our results with those in Refs.~\cite{frits:1998,frits:1999}. 
We emphasise
that our work differs from the work of Ref.~\cite{frits:1998} in that
we include the exact EW ${\cal O}(\alpha)$ corrections with
YFS exponentiation in the production process and we actually calculate
the effects of the FSR in the $W$-pair 
production and decay process
at LEP2/NLC energies; in Ref.~\cite{frits:1998}, only the process
$\nu_\mu \bar\nu_\mu\rightarrow ZZ\rightarrow e^+ e^-+\nu_\tau \bar\nu_\tau$
is actually calculated, and a heuristic argument is used to 
estimate the corresponding results for final-state $W$-decay radiation.
Thus, our calculations will also be a comment on the accuracy of these 
heuristic arguments in the presence of the YFS exponentiated exact 
${\cal O}(\alpha)$ corrections to the  $W$-pair
production process. Our work differs from that
in Ref.~\cite{frits:1999} in that we include the YFS exponentiation
of the exact ${\cal O}(\alpha)$ production process in the $W$-pair
intermediate state and the ${\cal O}(\alpha)^2$ LL FSR whereas,
in Ref.~\cite{frits:1999}, the exact ${\cal O}(\alpha)$
correction to the production and decay processes for the $W$-pair
in the leading pole approximation is calculated without exponentiation. 
The leading pole approximation treatment of the attendant 
non-factorizable corrections in Refs.~\cite{mel:1996,fritsnf:1997,ditm:1998}
is also retained. The latter non-factorizable corrections
have been shown~\cite{mel:1996,fritsnf:1997,ditm:1998} to be small
and, as we explain in Ref.~\cite{yfsww3:1998}, when one works up to but not
including ${\cal O}(\frac{\alpha}{\pi}\frac{\Gamma_W}{M_W})$ as we do,
such effects may be dropped; which we do. Thus, although we start
our calculation in Ref.~\cite{yfsww3:1998} 
in the fermion-loop scheme~\cite{been2:1996},
when we focus on the ${\cal O}(\alpha)$ EW correction, we go to the
leading pole part of the respective production amplitude. We also make the
approximation of using on-shell residues for this double pole part,
with which we then approximate the corresponding ${\cal O}(\alpha)$
EW correction. In our Monte Carlo event generator approach,
we stress that the full off-shell phase space is always retained here.
We improve our result by using the complete on-shell
residues for EW corrections
rather than their on-shell fermion-loop scheme representatives.
Indeed, for the QED bremsstrahlung correction we stress that,
since the real photon has $k^2=0$, the corresponding running
charge is the usual one. It can thus be shown that, in ${\cal O}(\alpha)$,
bremsstrahlung residues of the on-shell fermion-loop scheme are equivalent
to those in the LPA; in both cases
all infrared singularities are properly cancelled and not only
the QED gauge invariance is preserved but also the full
$SU_{2L}\times U_1$ gauge invariance~\cite{yfsww2:1996,yfsww3:1998}. 
For this reason, 
in order ${\cal O}(\alpha)$, in our final result, any reference to the
fermion-loop scheme is purely pedagogical. What we arrive at is
precisely the LPA, with full on-shell residues for the respective
double pole approximation.
Indeed, as the YFS expansion is not generally familiar, if one looks at
Eqs. (1) and (2) in Ref.~\cite{yfsww3:1998}, which give the 
on-shell ${\cal O}(\alpha)$ contributions to the YFS residuals
$\bar\beta_0$ and $\bar\beta_1$, respectively, for the production
process in the LPA, one may think that the lowest-order contribution
to $\bar\beta_0$, $\bar\beta^{(0)}_0$ in 
the notation of Ref.~\cite{yfsww3:1998}, is {\em not} required either to be 
evaluated at the corresponding on-shell point as well. 
However, the right-hand side
of Eq.~(2) in Ref.~\cite{yfsww3:1998}, for example, involves the
subtraction, from the corresponding on-shell ${\cal O}(\alpha)$ 
bremsstrahlung cross section, 
of the product of the YFS real emission infrared 
function $\tilde S$~\cite{yfs:1961,yfsww2:1996} by the on-shell 
lowest order Born cross section;
{\em we need to stress that the} \underline{YFS theory} {\em then forces
the contribution to $\bar\beta_0$ corresponding 
to this respective lowest-order Born cross section,
$\bar\beta^{(0)}_0$, to be evaluated at the
on-shell point as well}. 
Thus, according to the YFS theory, Eqs.~(1) and (2) in Ref.~\cite{yfsww3:1998}
are entirely equivalent to results in Refs.~\cite{frits:1999} for the 
production process, for the contributions up to and including terms 
${\cal O}(\alpha)$.
%
%
As can also be seen from the results in Refs.~\cite{frits:1999},
these approximations are valid up to but not including
${\cal O}(\frac{\alpha}{\pi}\frac{\Gamma_W}{M_W})$. We then apply the
YFS Monte Carlo methods of two of us (S.J. and B.F.L.W.)~\cite{jw:1988},
as extended to spin~1 particles in Ref.~\cite{yfsww2:1996},
to arrive at the respective
exact ${\cal O}(\alpha)_{prod}$ YFS exponentiated results 
realized in YFSWW3-1.11. Hence, we stress that, as far as the
${\cal O}(\alpha)$ 
correction to the production process under study
is concerned, the results in Refs.~\cite{frits:1999,yfsww3:1998} 
should be equivalent, in view of the
many cross checks carried out by the authors of Refs.~\cite{ew1,ew2,ew3}
on the two corresponding electroweak on-shell calculations used therein.

More precisely, starting from the calculations in the program
YFSWW3-1.11 in Ref.~\cite{yfsww3:1998}, which feature the
exact ${\cal O}(\alpha)_{prod}$ YFS exponentiated results
for the process 
$e^+e^-\rightarrow W^+W^-+n(\gamma)\rightarrow 4f + n(\gamma)$,
we have interfaced the outgoing final state
to the program PHOTOS~\cite{photos:1994}. The latter uses the structure
function evolution equation for QED~\cite{strfn} to generate up to 
two final-state decay photons for each $W$ according to the 
respective LL probabilities to radiate; here the corresponding
angular distributions of the decay photons are all generated
in accordance with this LL approximation as described in
Ref.~\cite{photos:1994}. The net probability of the respective event
is unchanged, i.e. the normalization of YFSWW3-1.11 is unaffected by this
interface, which will be described in more detail elsewhere~\cite{elsewh}.
We refer to the version of YFSWW3 that contains this final-state radiative
interface to PHOTOS as YFSWW3-1.12 and it is available from the
authors~\cite{yfsww3-1.12}.
In what follows, we present some sample Monte Carlo data from
YFSWW3-1.12 to look into the possible role of FSR in the presence
of the ${\cal O}(\alpha)$ EW corrections. For definiteness, 
we focus here on the current LEP2 CMS energy of $190$~GeV and on the
SM couplings. The complete discussion of both LEP2 and NLC energies
with the illustration of anomalous couplings will 
appear elsewhere~\cite{elsewh}. 

\let\sstl=\scriptscriptstyle
\def\Was{W\c as}
\def\Order#1{${\cal O}(#1$)}
\def\Ordpr#1{${\cal O}(#1)_{prag}$}
\def\bbe{\bar{\beta}}
\def\tbe{\tilde{\beta}}
\def\tal{\tilde{\alpha}}
\def\tom{\tilde{\omega}}
\def\half{ {1\over 2} }
\def\alf1{ {\alpha\over\pi} }

\def\Oaz{${\cal O}(\alpha^0)$}
\def\Oaf{${\cal O}(\alpha^1)$}
\def\Oas{${\cal O}(\alpha^2)$}
\begin{table}[!ht]

\centering
\begin{tabular}{||c|c|c|c|c||}
\hline\hline
 $E_{CM}\,{\rm [GeV]}$  & Calculation & FSR & 
 CUT &$\langle M_W\rangle \,{\rm [GeV]}$\\
\hline 
~&~&$W^-\rightarrow e^-\bar\nu_e$& ~&\\ 
\hline
$190$ & {\it Born} &--  & $ BARE$ 
      & $80.253\pm 0.008$ \\
      & {\it EW-ex} & $NO$ & $BARE$ 
      & $80.146\pm 0.036$ \\
      & {\it No~EW} & $NO$ & $BARE$ 
      & $80.142\pm 0.036$ \\      
      & {\it No~EW} & $YES$ & $BARE$ 
      & $78.614\pm 0.035$ \\
      & {\it EW-ap} & $YES$ & $BARE$
      & $78.613\pm 0.035$ \\
      & {\it EW-ex} & $YES$ & $BARE$ 
      & $78.618\pm 0.035$ \\
      & {\it No~EW} & $YES$ & $CALO$ 
      & $79.727\pm 0.036$ \\
      & {\it EW-ap} & $YES$ & $CALO$
      & $79.725\pm 0.036$ \\
      & {\it EW-ex} & $YES$ & $CALO$ 
      & $79.731\pm 0.036$ \\
\hline 
~ &~&$W^-\rightarrow \mu\bar\nu_\mu$&~& \\ 
\hline
$190$ & {\it Born} &--  & $ BARE$ 
      & $80.253\pm 0.008$ \\
      & {\it EW-ex} & $NO$ & $BARE$ 
      & $80.146\pm 0.036$ \\
      & {\it No~EW} & $NO$ & $BARE$ 
      & $80.142\pm 0.036$ \\
      & {\it No~EW} & $YES$ & $BARE$ 
      & $79.374\pm 0.036$ \\
      & {\it EW-ap} & $YES$ & $BARE$
      & $79.373\pm 0.036$ \\
      & {\it EW-ex} & $YES$ & $BARE$ 
      & $79.378\pm 0.036$ \\
      & {\it No~EW} & $YES$ & $CALO$ 
      & $79.725\pm 0.036$ \\
      & {\it EW-ap} & $YES$ & $CALO$
      & $79.724\pm 0.036$ \\
      & {\it EW-ex} & $YES$ & $CALO$ 
      & $79.730\pm 0.036$ \\
\hline\hline
\end{tabular}

\caption{\small\sf The results of the $125\times10^6$ 
statistics samples (weighted events)
(except for $Born$, where the sample is $540\times10^6$ of such events)
from YFSWW3-1.12 for the average value of $M_W$ as computed
with the levels of radiative corrections as indicated
for both bare and calorimetric treatments of the final lepton.
See the text for more details.
}
\label{tab:xsec}
\end{table}

Specifically, in Figs.~1--8, we show the results obtained with
YFSWW3-1.12 on the processes $e^+e^-\rightarrow W^+W^-+n(\gamma)
\rightarrow \bar c s+\ell \bar\nu_{\ell}$, $\ell = e,~\mu$,
for the cosine of the $W$ production angle distribution in the CM
(LAB) system, for the $W$ mass distribution, with both 
``bare'' and ``calorimetric''
definitions of that mass, for the CMS lepton final energy distribution,
for both calorimetric and bare definitions of that energy, and
for the corresponding distributions of the cosine of the  
lepton decay angle in the
$W$ rest frame.  
We note the following
properties of these results. First, concerning the $W$ mass distributions
in Figs.~1 and 5, we see that the respective average values of $M_W$
are as given in Table~1. There, {\it EW-ex} denotes the exact
${\cal O}(\alpha)_{prod}$ calculation of EW corrections~\cite{yfsww3:1998};
{\it EW-ap} denotes the approximate treatment of these EW corrections as given
in Ref.~\cite{sigap}; {\it No~EW} denotes that the EW corrections
other than the ones coming from 
LL (${\cal O}(\alpha^2)$) initial-state radiation 
are turned off. The calorimetric results are all closer to their respective
{\it NO FSR} analogues than are the bare results, as expected.
The effects of the FSR for the muon case are all respectively smaller than
the corresponding results for the electron case, again as expected
because of the smaller radiation probability for the muon. The size of the
shift of $\langle M_W\rangle$ is generally consistent with the discussion in
Ref.~\cite{frits:1998}, which deals with primarily the line shape 
(peak position and height); in detail we see that, 
in the presence of the FSR, at the level of our statistical errors,
for an average quantity such as $\langle M_W\rangle$, all three 
calculations in the table are sufficient, as expected.
With regard to the guesstimates made in Ref.~\cite{frits:1998}
concerning the peak reduction and the peak position shift,
we see from the $BARE$ curves in Fig.~1 that our result
of $13.5\%$ for the peak reduction in the $e^-$ case 
(comparing the {\it EW-ex} curves with and without FSR) is in good agreement
with the $14\%$ guesstimate of Ref.~\cite{frits:1998}
and with the $14.4\%$ found in the recent ${\cal O}(\alpha)$ on-shell
LPA results in Ref.~\cite{frits:1999}. The $\sim-57$~MeV
guesstimated in Ref.~\cite{frits:1998} for the 
corresponding peak position shift in the $e^-$ case was recently
updated to $-77$~MeV in Ref.~\cite{frits:1999};
for the $\mu$ case, the updated
expectation from Ref.~\cite{frits:1999} for the peak position shift
is $-39$~MeV. For completeness, we note that the size of the
peak reduction effect in the $\mu$ case has been found to be
$\sim 8\%$ in Ref.~\cite{frits:1999} whereas in Fig.~5 we find
$7.6\%$, again showing good agreement
between our results and those in Ref.~\cite{frits:1999}.
Indeed, to compare our results for the peak position shift with
those just cited from Ref.~\cite{frits:1999}, we have performed
Breit--Wigner fits to our line shapes in Figs.~1 and 5 with the
values, both fixed and floating, of the $W$ width. The results of our
fits are shown in Table~2.
\begin{table}[!ht]
\centering
{\footnotesize
\begin{tabular}{||l|c|cc|cc|cc||}
\hline\hline
 \multicolumn{8}{||c||}{ $M_W$ or $M_W$/$\Gamma_W$ [GeV]} \\
 \hline\hline
 \multicolumn{8}{||c||}{$W^-\rightarrow e^-\bar\nu_e$} \\
 \hline
             & $M$-range & 
 \multicolumn{2}{|c}{\it No FSR} & \multicolumn{2}{|c}{\it FSR-BARE} & 
 \multicolumn{2}{|c||}{\it FSR-CALO}\\
\cline{3-8} 
 & [GeV]&$\Gamma_W$-fix & $\Gamma_W$-fit & 
         $\Gamma_W$-fix & $\Gamma_W$-fit & 
         $\Gamma_W$-fix & $\Gamma_W$-fit \\
\hline
 {\it Born} &78 -- 82 &
 $80.240$ & $80.240$/$2.0413$ & & & & \\
          &76 -- 84 &
 $80.239$ & $80.239$/$2.0376$ & & & & \\
\hline
 {\it No EW} &78 -- 82 & 
 $80.231$ & $80.231$/$2.0442$ & 
 $80.166$ & $80.168$/$2.2105$ &  
 $80.216$ & $80.217$/$2.0831$   \\ 
             &76 -- 84 & 
 $80.227$ & $80.227$/$2.0372$ & 
 $80.142$ & $80.135$/$2.2547$ &  
 $80.207$ & $80.207$/$2.0892$   \\ 
\hline
 {\it EW-ap} &78 -- 82 & 
          &                   & 
 $80.166$ & $80.168$/$2.2105$ &  
 $80.216$ & $80.217$/$2.0832$   \\ 
             &76 -- 84 & 
          &                   & 
 $80.142$ & $80.134$/$2.2547$ &  
 $80.207$ & $80.207$/$2.0892$   \\ 
\hline
 {\it EW-ex} &78 -- 82 & 
 $80.231$ & $80.231$/$2.0443$ &                    
 $80.166$ & $80.168$/$2.2105$ &  
 $80.216$ & $80.217$/$2.0832$   \\ 
             &76 -- 84 & 
 $80.227$ & $80.227$/$2.0372$ & 
 $80.142$ & $80.134$/$2.2547$ &  
 $80.207$ & $80.207$/$2.0892$   \\ 
\hline

 \multicolumn{8}{||c||}{$W^-\rightarrow \mu^-\bar\nu_{\mu}$} \\
\hline
 {\it Born} &78 -- 82 &
 $80.241$ & $80.241$/$2.0308$ & & & & \\
          &76 -- 84 &
 $80.250$ & $80.250$/$2.0295$ & & & & \\
\hline
 {\it No EW} &78 -- 82 & 
 $80.232$ & $80.232$/$2.0342$ & 
 $80.198$ & $80.199$/$2.1196$ &  
 $80.217$ & $80.218$/$2.0731$   \\ 
             &76 -- 84 & 
 $80.238$ & $80.238$/$2.0307$ & 
 $80.192$ & $80.190$/$2.1481$ &  
 $80.217$ & $80.217$/$2.0845$   \\ 
\hline
 {\it EW-ap} &78 -- 82 & 
          &                   & 
 $80.198$ & $80.199$/$2.1196$ &  
 $80.217$ & $80.218$/$2.0731$   \\ 
             &76 -- 84 & 
          &                   & 
 $80.192$ & $80.190$/$2.1481$ &  
 $80.217$ & $80.217$/$2.0845$   \\ 
\hline
 {\it EW-ex} &78 -- 82 & 
 $80.232$ & $80.232$/$2.0343$ &                    
 $80.198$ & $80.199$/$2.1196$ &  
 $80.217$ & $80.218$/$2.0731$   \\ 
             &76 -- 84 & 
 $80.238$ & $80.238$/$2.0307$ & 
 $80.192$ & $80.190$/$2.1481$ &  
 $80.217$ & $80.217$/$2.0845$   \\ 
\hline\hline
\end{tabular}
}
\caption{\small\sf The results of the Breit--Wigner line shape fit
to the YFSWW3-1.12 MC sample for the $W^-$ invariant mass distribution
at $E_{CMS} = 190$~GeV. The input values of the $W$ mass and width were: 
$M_W=80.23$~GeV and $\Gamma_W=2.03367033$~GeV (this value was used in 
the $\Gamma_W$-fix fit). The fits were performed for two $W$ invariant
mass $M$ ranges -- as indicated in the table.
See the text for more details.
}
\label{tab2:xsec}
\end{table}
%
For comparison, the fits are done for two
different mass intervals, from $78$~GeV to $82$~GeV, and
from $76$~GeV to $84$~GeV, to illustrate the role of the wings of the resonance
in the fits. From these results we find that the {\it BARE}
peak position shifts are estimated using the narrow fit range
as $80.168 - 80.240=-72$~MeV and $80.199 - 80.241=-42$~MeV for the
$e$ and $\mu$ cases, respectively.
We also computed the shift in the average 
invariant mass $\langle M_W\rangle$ of the $W$ in the narrow range 
from $78$~GeV to $82$~GeV as another estimate of the peak position shift
for the {\it BARE} trigger and we found $-81.5\pm1.4$~MeV and
$-43.9\pm0.9$~MeV for the $e$ and $\mu$ cases, respectively.
Thus, both sets of estimators of the peak position
shifts are in reasonable agreement with
the results given in Ref.~\cite{frits:1999}%
\footnote{ The fit mass shift and
the peak position shift approach one another as the fit range 
approaches a zero size interval around the peak;
a similar remark applies to the shift in the
average mass relative to the range over which it is taken
around the peak.}; in this connection, we recall the slight difference in
beam energy between our studies ($95$~GeV) and those in Ref.~\cite{frits:1999}
($92$~GeV).
Moreover, we see
in Table~2 the same pattern of results as we see in Table~1: the 
FSR effects for the $e$ case are more pronounced than those for the
$\mu$ case; the calorimetric acceptance reduces the size of the FSR effects;
the results are not very sensitive to the EW correction
to the production process. 
If we compare the predictions
with and without FSR for the {\it EW-ex} and {\it no~EW} cases
we get a measure of the modulation of the FSR on the EW correction.
From the curves in our Figs.~1 and 5 and the respective plots of
the $\delta_{RAD}$ as defined in the figures we see that this
modulation is as expected.
Concerning the cosine of the $W$ production angle distributions, 
we see the interplay
of the exact EW corrections on the one hand and the FSR on the other.
Further, we see that the approximate EW corrections of Ref.~\cite{sigap}, 
while a definite improvement over the no EW corrections at all,
are not sufficient to describe this interplay at the level of 
$0.5$--$1.0\%$. Similar remarks hold for the lepton energy distribution in the 
CM system, although the corresponding insufficiency is reduced to the level
of $\sim 0.3\%$ for the $BARE$ case, for example for electrons.
Concerning the distributions of the cosine of the lepton decay angle
in the $W$ rest frame, we again see the importance of including both
the EW corrections and the FSR in Figs.~4 and~8, for the electron
and the muon, respectively.
In all cases, the results for the muon, particularly the $BARE$ results,
are less affected by the FSR than are the corresponding results
for the electron, as expected. We stress 
that our results in Figs. 1--8 are generally
consistent with those in Ref.~\cite{frits:1999} keeping in mind
that we treat the ${\cal O}(\alpha^2)$ LL FSR and the YFS exponentiated
on-shell exact ${\cal O}(\alpha)_{prod}$ production process,
whereas Ref.~\cite{frits:1999} treats only ${\cal O}(\alpha)$
corrections in our LPA, in which only on-shell residues are used.
Indeed, in addition to the agreements already cited,
we call the reader's attention to the normalization correction in 
Fig.~9 of Ref.~\cite{frits:1999}: at the CMS energy of $\sqrt{s}=190$~GeV,
it is $-11\%$, in very good agreement with our result in 
Ref.~\cite{yfsww3:1998}, which is $(1+\delta_{prod})(\rho_w)^2-1\cong-11.1\%$,
for the latter result, we have used Table~2 in Ref.~\cite{yfsww3:1998} for the
relative correction $\delta_{prod}=-9.9\%$ to the production
process, and the result in Ref.~\cite{dima:1986} for the
${\cal O}(\alpha)$ correction to the leptonic partial width
$\rho_w-1\cong -0.686\%$. In addition, we can note that, for
the case of the $\tau\bar\nu_\tau$ decay channel, our results
are also consistent with those in Ref.~\cite{frits:1999}
for the peak position shift and peak reduction effects.
In view of our higher-order corrections, we find quite reasonable all 
the agreements noted here. A more detailed discussion
of such comparisons will appear~\cite{elsewh}. We stress that
we have arrived at our results through a MC event generator realization
of our calculation, in which realistic, finite $p_T$, $n(\gamma)$
radiation is incorporated in the production process on an event-by-event
basis, whereas the results in Ref.~\cite{frits:1999} are
all semi-analytical. This enhances the significance of the
general agreement of our results where they do overlap.
\par

The issue of whether the calorimetric
results are more realistic than the bare ones appears to depend
on whether one is talking about the muon or the 
electron%
\footnote{T. Kawamoto, private communication, 1998.}.
For the electron, it is very difficult
to separate the soft photons with energy $\lesssim \Gamma_W$
that are responsible for the FSR effects of the $W$ line shape as
discussed already in Refs.~\cite{frits:1998,frits:1999}; they are just a part
of the electromagnetic calorimeter response in general, which is used
to measure the electron energy. For the muon, the energy is usually
measured by a muon chamber in which, in general, these soft photons
are not present. Thus, for the electron, our calorimetric results
are more realistic; for the muon, it is the other way around.
In either case, we see that precision $W$-pair production and decay
studies need to take the interplay between the FSR and the EW
corrections into account so as to obtain the most precise tests
of the SM; our calculations in YFSWW3-1.12 offer
an avenue to achieve that goal.

\vspace{7mm}
\noindent 
{\large\bf  Acknowledgements}

Two of us (S.J. and B.F.L.W.) acknowledge the
kind hospitality of Prof. A. De R\'ujula and the CERN Theory 
Division while this work was being completed. 
Two of us (B.F.L.W. and W.P.) 
acknowledge the support of Prof.~D.~Schlatter
and the ALEPH Collaboration
in the final stages of this work. 
One of us (Z.W.) acknowledges the support of the L3 group of ETH Zurich 
during the time this work was performed.


\begin{figure}
\begin{center}
\epsfig{file=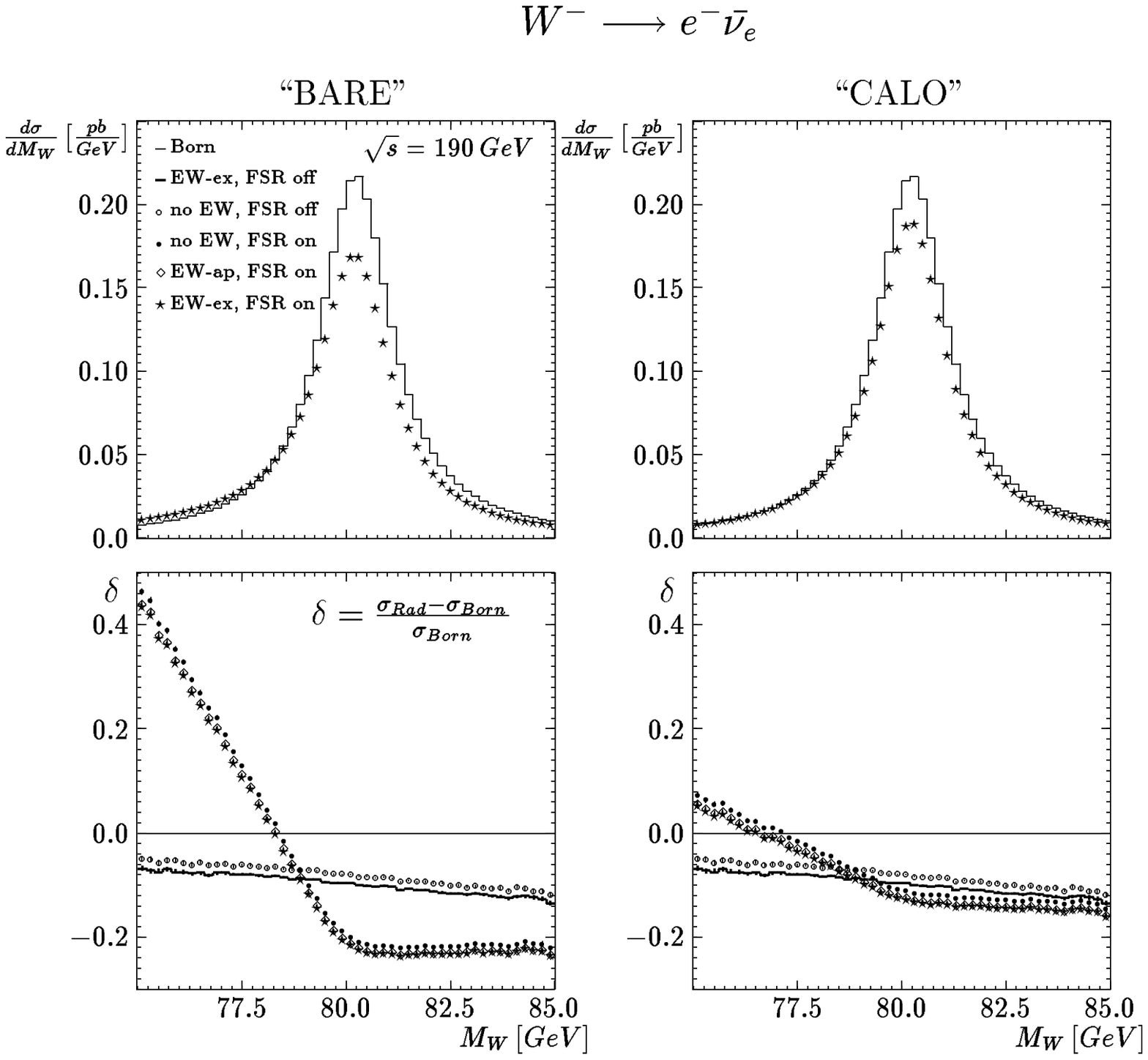,height=15cm}
\end{center}
\caption{\small\sf 
The invariant mass distributions of $W^-$ reconstructed from its decay
products, \protect$e^- \bar{\nu_e}$, four-momenta. 
In the left pictures the electron
is treated exclusively (`bare' electron), while in the right pictures
it is treated calorimetrically (`dressed' electron -- its four-momentum
is combined with four-momenta of all photons emitted within an angle of
$5^\circ$ around its direction). The input values are $M_W = 80.23$~GeV, 
~$\Gamma_W= 2.034$~GeV.
} 
\label{fig:el2}
\end{figure}
\noindent
\begin{figure}
\begin{center}
\epsfig{file=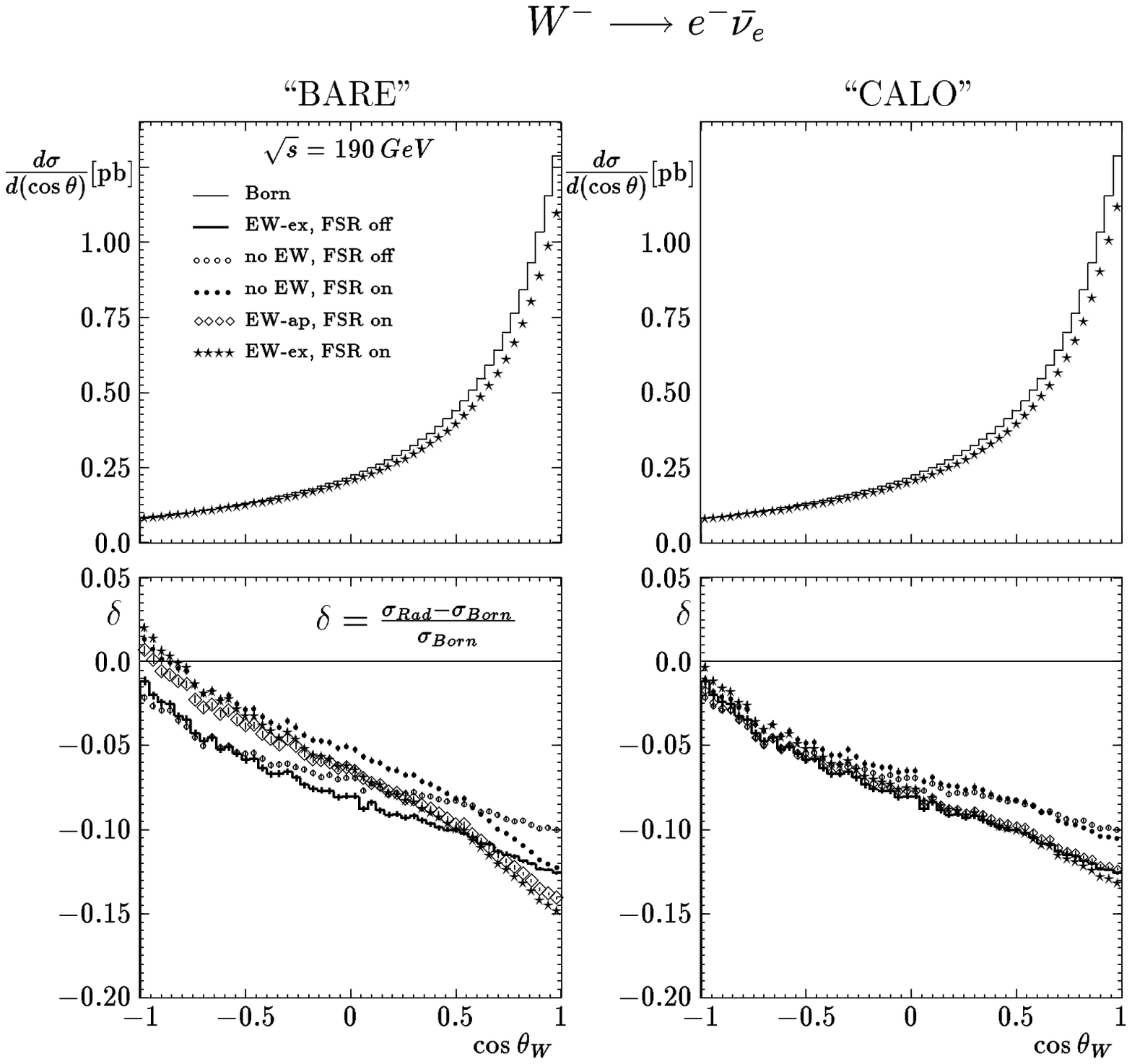,height=15cm}
\end{center}
\caption{\small\sf
The angular distributions of $W^-$ reconstructed from its decay
products, \protect$e^- \bar{\nu_e}$, four-momenta. 
In the left pictures the electron
is treated exclusively (`bare' electron), while in the right pictures
it is treated calorimetrically as defined in Fig.~1.
}
\label{fig:el1}
\end{figure}
\noindent
\begin{figure}
\begin{center}
\epsfig{file=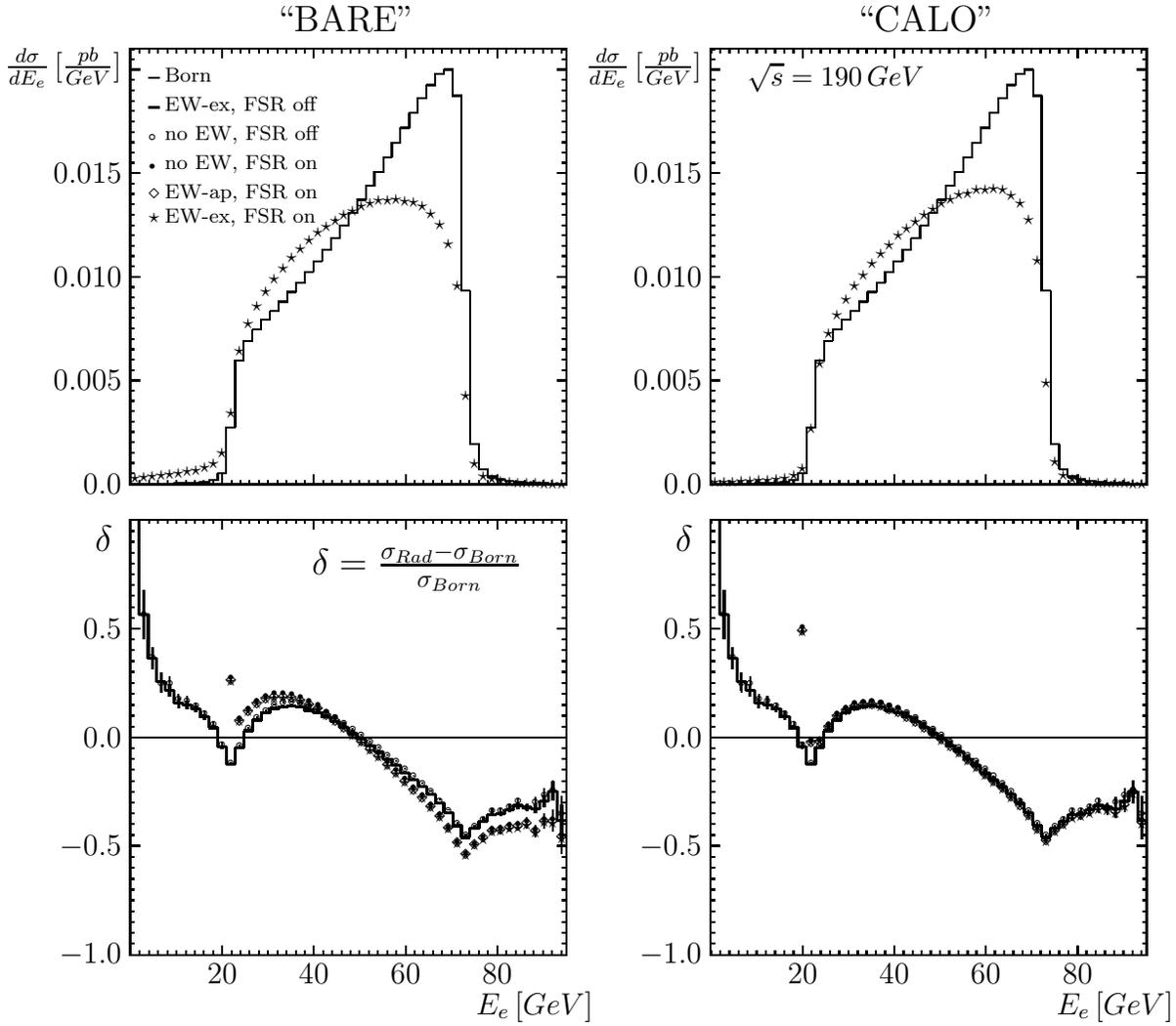,height=15cm}
\end{center}
\caption{\small\sf
The distributions of the final state electron energy in the LAB frame. 
In the left pictures the electron
is treated exclusively (`bare' electron), while in the right pictures
it is treated calorimetrically as defined in Fig.~1.
} 
\label{fig:el3}
\end{figure}
\noindent
\begin{figure}
\begin{center}
\epsfig{file=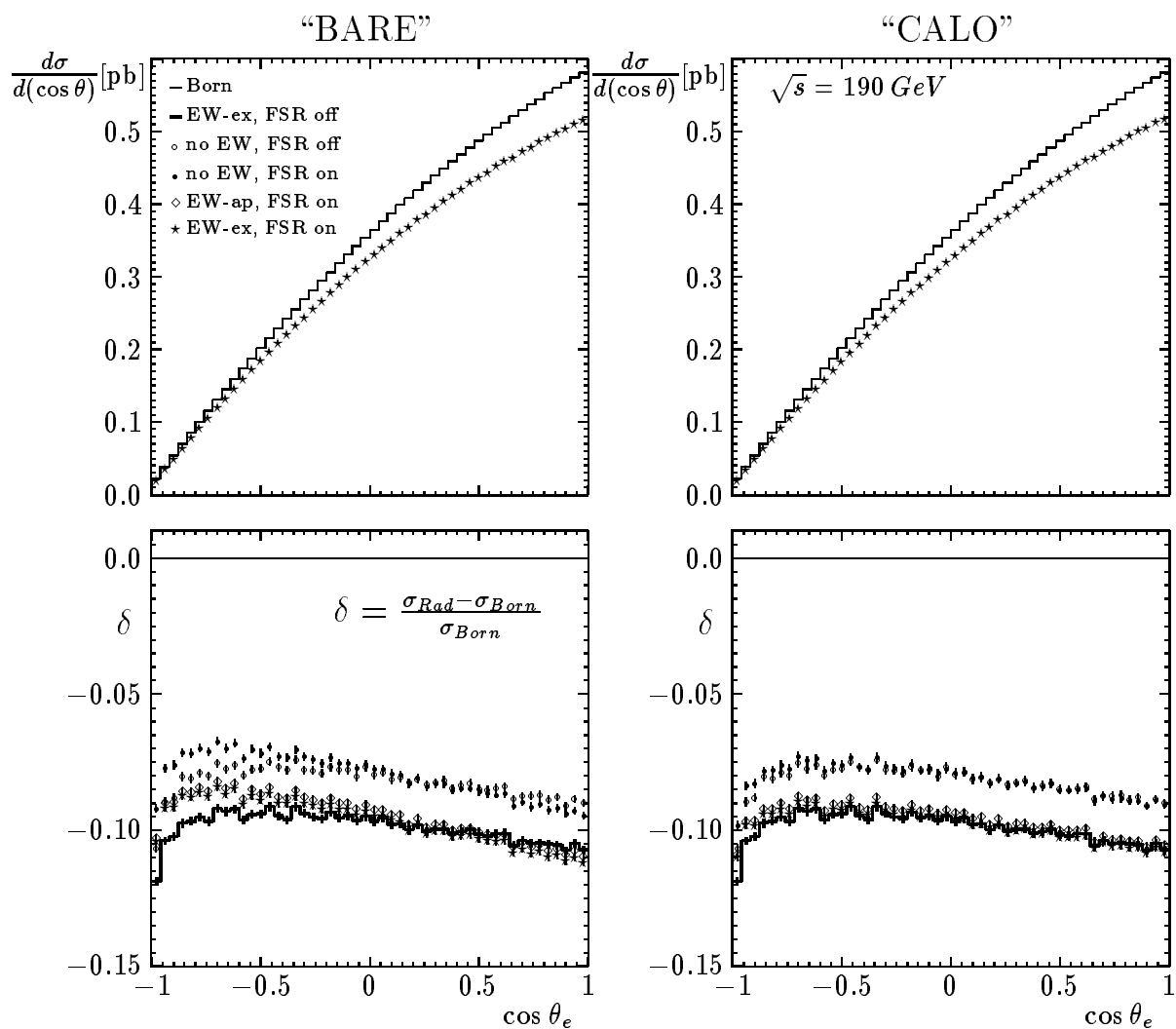,height=15cm}
\end{center}
\caption{\small\sf
The distributions of the electron decay angle's cosine in the $W^-$ rest frame.
In the left pictures the electron
is treated exclusively (`bare' electron), while in the right pictures
it is treated calorimetrically as defined in Fig.~1.
}
\label{fig:el4}
\end{figure}
\noindent

\begin{figure}
\begin{center}
\epsfig{file=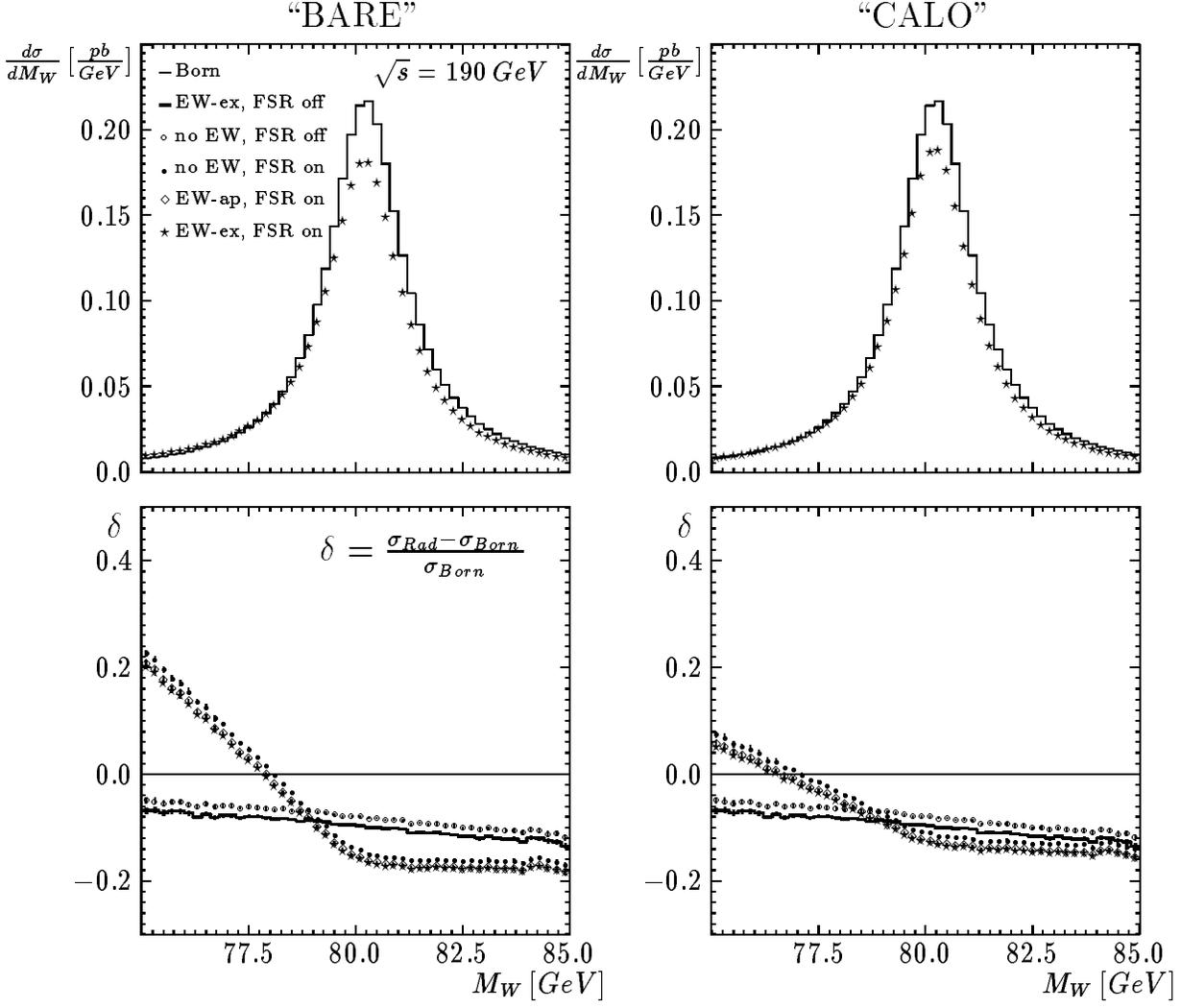,height=15cm}
\end{center}
\caption{\small\sf
The invariant mass distributions of $W^-$  reconstructed from its decay
products, \protect$\mu^- \bar{\nu_{\mu}}$, four-momenta. 
In the left pictures the muon
is treated exclusively (`bare' muon), while in the right pictures
it is treated calorimetrically as defined in Fig.~1. 
The input values are $M_W = 80.23$~GeV,~$\Gamma_W = 2.034$~GeV.
} 
\label{fig:mu2}
\end{figure}
\noindent

\begin{figure}
\begin{center}
\epsfig{file=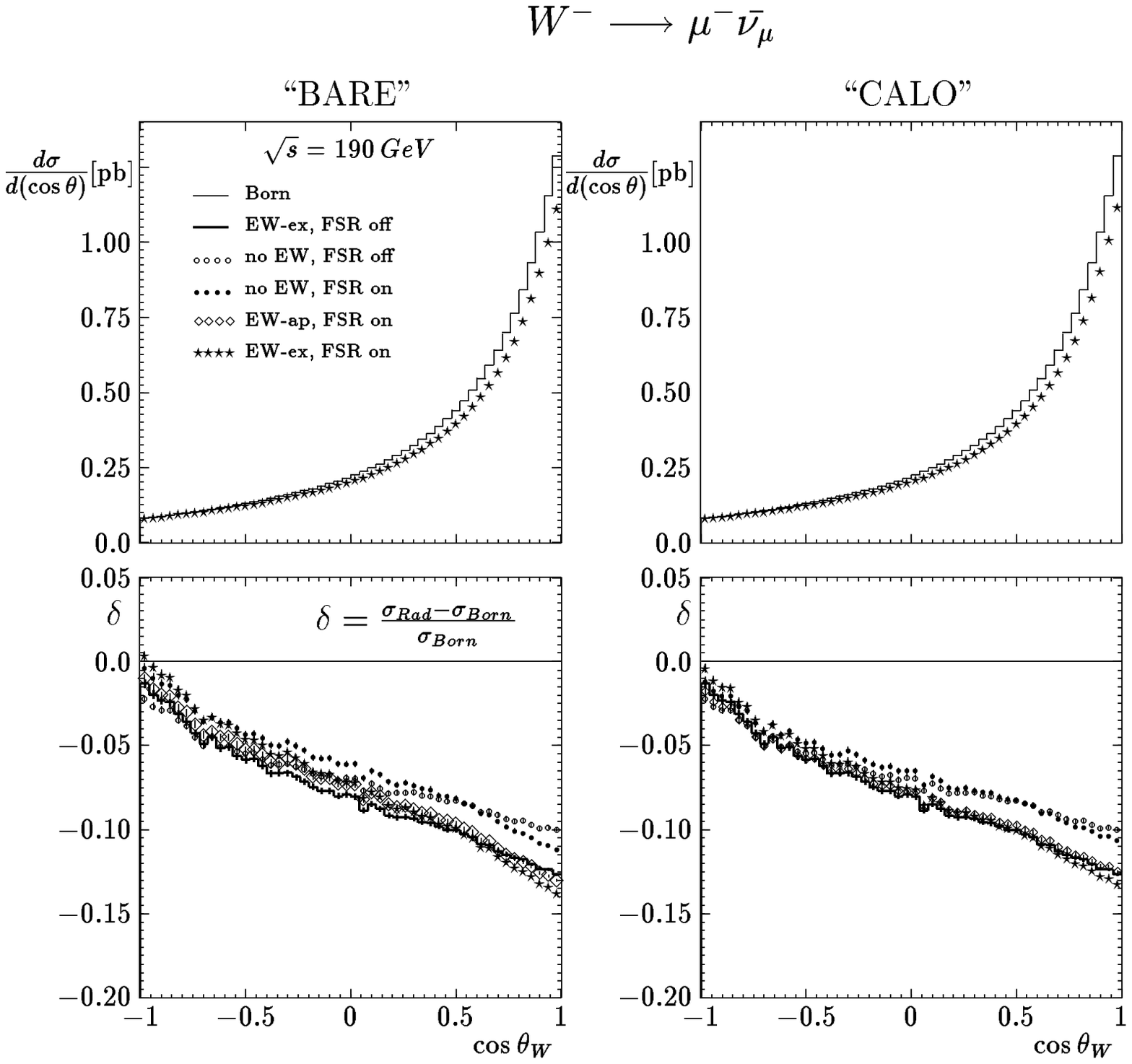,height=15cm}
\end{center}
\caption{\small\sf
The angular distributions of $W^-$ reconstructed from its decay
products, \protect$\mu^- \bar{\nu_{\mu}}$, four-momenta. 
In the left pictures the muon
is treated exclusively (`bare' muon), while in the right pictures
it is treated calorimetrically as defined in Fig.~1.
}
\label{fig:mu1}
\end{figure}
\noindent
\begin{figure}
\begin{center}
\epsfig{file=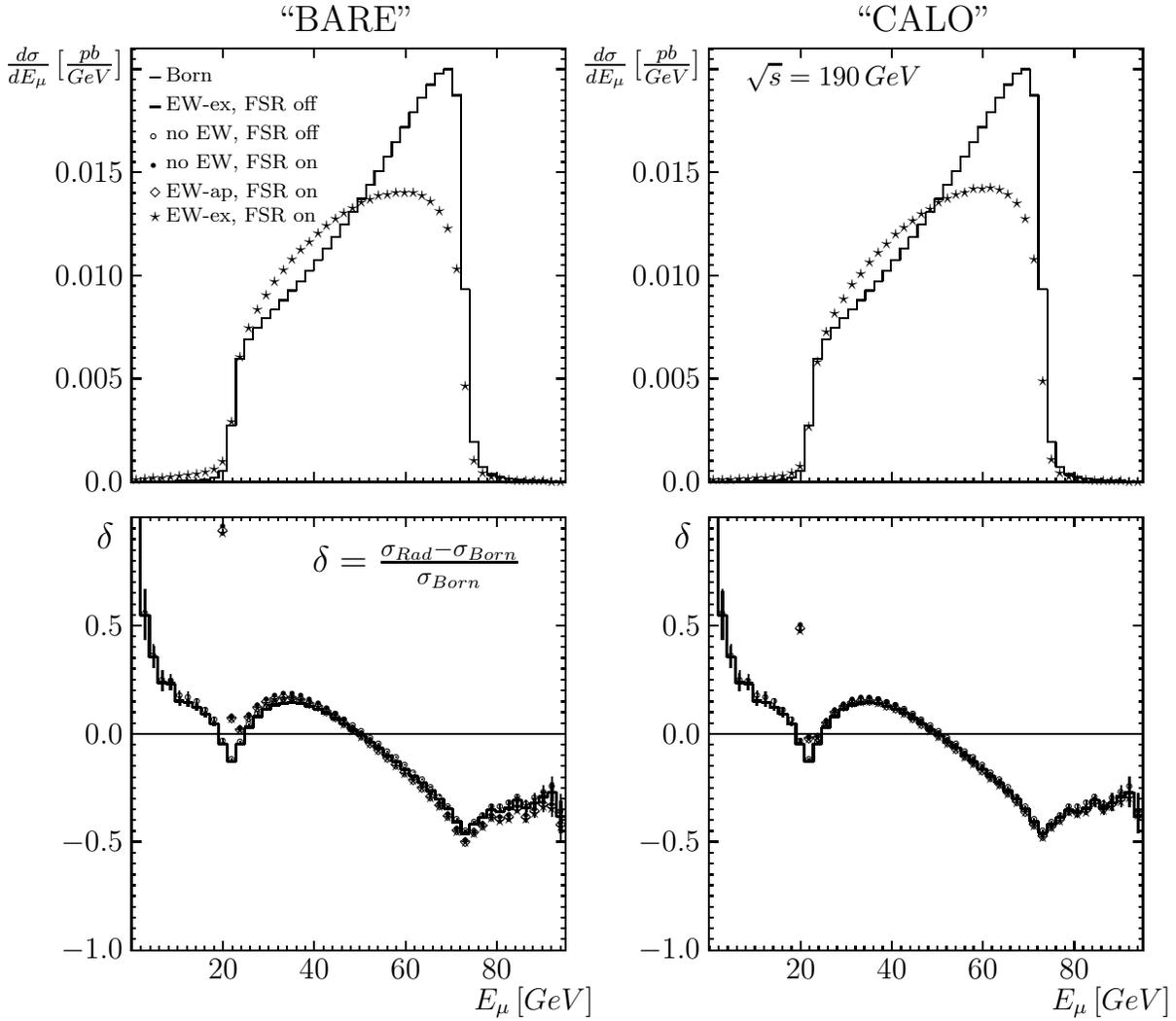,height=15cm}
\end{center}
\caption{\small\sf
The distributions of the final state muon energy in the LAB frame. 
In the left pictures the muon
is treated exclusively (`bare' muon), while in the right pictures
it is treated calorimetrically as defined in Fig.~1.
} 
\label{fig:mu3}
\end{figure}
\noindent
\begin{figure}
\begin{center}
\epsfig{file=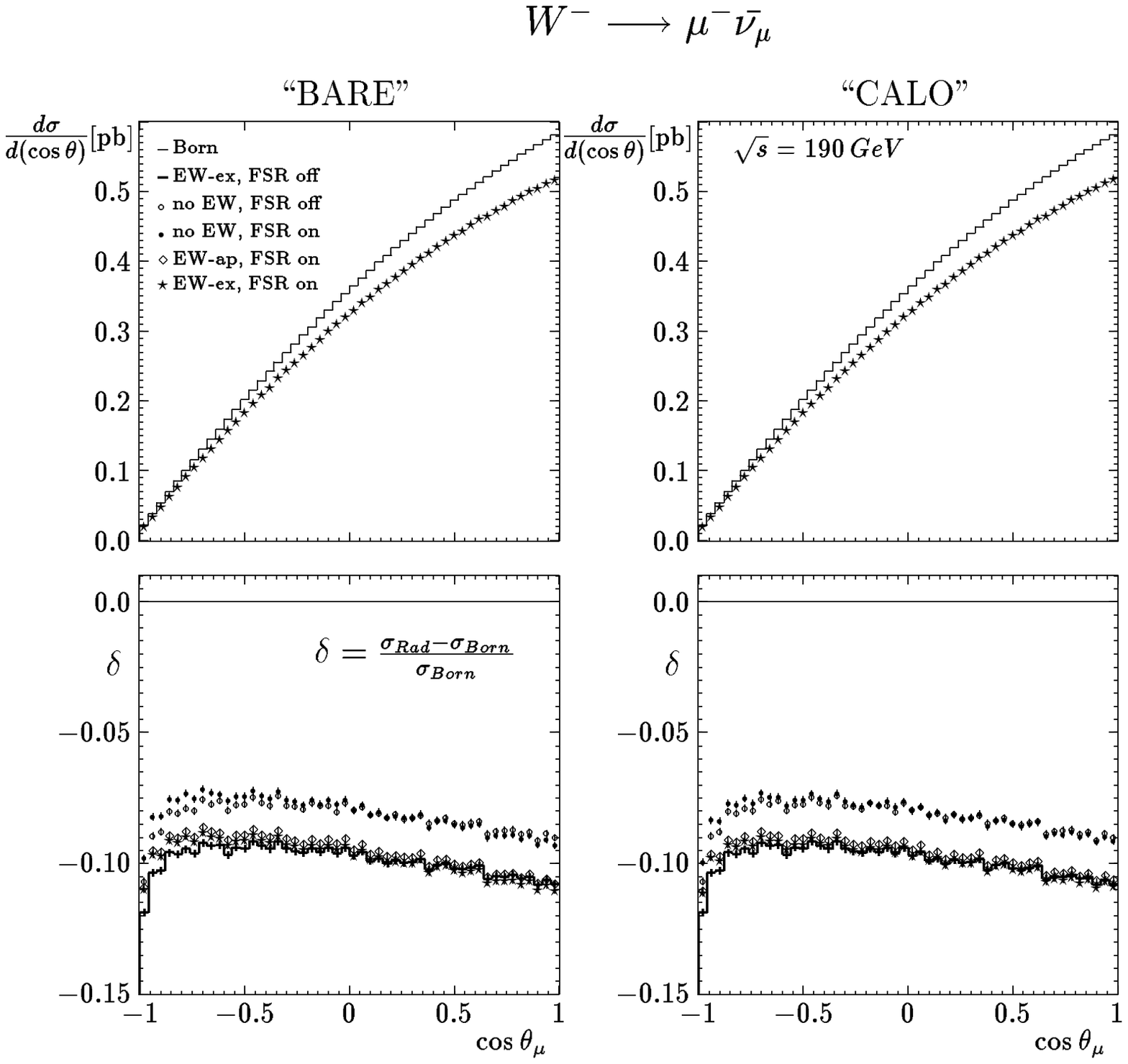,height=15cm}
\end{center}
\caption{\small\sf
The distributions of the muon decay angle's cosine in the $W^-$ rest frame.
In the left pictures the muon
is treated exclusively (`bare' muon), while in the right pictures
it is treated calorimetrically as defined in Fig.~1.
}
\label{fig:mu4}
\end{figure}


\begin{thebibliography}{99}
%
%
\bibitem{lep2ybk:1996}
W. Beenakker et al.,
{\em WW Cross-Sections and Distributions}, 
in {\em Physics at LEP2},
edited by G. Altarelli, T. Sj\"ostrand and F. Zwirner 
(CERN 96-01, Geneva, 1996), Vol.~1, p.~79.
%
\bibitem{frits:1998}
W.~Beenakker, F. A. Berends and A.~P.~Chapovsky, preprint hep-ph/9805327, 
1998; Phys. Lett. {\bf B435} (1998) 233.
%
\bibitem{frits:1999}
W.~Beenakker, F. A. Berends and A.~P.~Chapovsky, preprint hep-ph/9811481, 
1998; preprint hep-ph/9902333, 1999.
\bibitem{yfsww2:1996}
S. Jadach, W. P{\l}aczek, M. Skrzypek and B.F.L. Ward,
Phys. Rev. {\bf D54} (1996) 5434.
%
\bibitem{yfsww3:1998}
S. Jadach, W. P{\l}aczek, M. Skrzypek, B.F.L. Ward and Z. W\c{a}s,
Phys. Lett. {\bf B417} (1998) 326.
%
\bibitem{yfs:1961}
D. R. Yennie, S. Frautschi and H. Suura, Ann. Phys. {\bf 13} (1961) 379.
%
\bibitem{photos:1994}
E. Barberio and Z. W\c{a}s, Comput. Phys. Commun. {\bf 79} (1994) 291
and references therein.
%
\bibitem{strfn}
See for example, G. Altarelli and G. Martinelli, in 
{\em Physics at LEP}, eds. J. Ellis and R. Peccei
(CERN 86-02, Geneva, 1986), Vol.~1, p.~47; and references therein.
%
\bibitem{elsewh}
S. Jadach {et al.}, to appear.
%
\bibitem{robin:1991}
R. G. Stuart, Phys. Lett. {\bf B262} (1991) 113; \\
A. Aeppli, G.J. van Oldenborgh and D. Wyler, Nucl. Phys. {\bf B428} (1994) 126.
%
\bibitem{mel:1996}
K. Melnikov and O. Yakovlev, Nucl. Phys. {\bf B471} (1996) 90.
%
\bibitem{fritsnf:1997}
W.~Beenakker, F. A. Berends and A.~P.~Chapovsky,
Phys. Lett. {\bf B411} (1997) 203; Nucl. Phys. {\bf B508} (1997) 17.
%
\bibitem{yfsww3-1.12}
S. Jadach {et al.}, {\em YFSWW3~1.12 MC Event Generator},
available from the authors at the WWW URL 
http://enigma.phys.utk.edu/pub/YFSWW.
%
\bibitem{ditm:1998}
A. Denner, S. Dittmaier and M. Roth, Nucl. Phys. {\bf B519} (1998) 39;
Phys. Lett. {\bf B429} (1998) 145.
%
\bibitem{been2:1996}
U. Baur and D. Zeppenfeld, Phys. Rev. Lett. {\bf 75} (1995) 1002; \\
W. Beenakker {\it et al.}, Nucl. Phys. {\bf B500} (1997) 255 
and references therein.
%
\bibitem{jw:1988}
S. Jadach and B.F.L. Ward, Phys. Rev. D{\bf 38} (1988) 2897 
and {\bf 40} (1989) 3582;
Comput. Phys. Commun. {\bf 56} (1990) 351.
%
\bibitem{ew1}
J. Fleischer,  F. Jegerlehner and M. Zra\l{}ek, 
Z. Phys. {\bf C42} (1989) 409;\\
M. Zra\l{}ek and K. Ko\l{}odziej, Phys. Rev. {\bf D43} (1991) 43; \\
J. Fleischer, K. Ko\l{}odziej and F. Jegerlehner, 
Phys. Rev. {\bf D47} (1993) 830; \\
J. Fleischer {\it et al.}, Comput. Phys. Commun. {\bf 85} (1995) 29 
and references therein.
%
\bibitem{ew2}
M. B\"ohm {\it et al.}, Nucl. Phys. {\bf B304} (1988) 463.
%
\bibitem{ew3}
W. Beenakker {\it et al.}, Phys. Lett. {\bf B258} (1991) 469;
Nucl. Phys. {\bf B367} (1991) 287.
%
\bibitem{sigap}
S. Dittmaier, M. B\"ohm and A. Denner, Nucl. Phys. {\bf B376} (1992) 29
and {\bf B391} (1993) 483 (E).
%
\bibitem{dima:1986}
D. Yu. Bardin, S. Riemann and T. Riemann, Z. Phys. {\bf C32} (1986) 121
and references therein.
%
%
%
%
%
%
%
%
%
%
%
%
%
%
%
%
%
%
%
%
%
%
%
%
%
%
%
%
%
%
%
%
%
%
%
%
%
%
%
%
%
%
%
%
%
%
\end{thebibliography}
\end{document}